\documentclass[twocolumn,amstex]{article}
\usepackage{graphicx}

\begin{document}


\title{An additional planet as a model for the Pleistocene Ice Age} 

\author{W. W\"olfli\footnote{Institute for Particle Physics, ETHZ
    H\"ong\-ger\-berg, CH-8093 Z\"urich, Switzerland (Prof.~emerit.); e-mail: woelfli@particle.phys.ethz.ch},
  W. Baltensperger\footnote{Centro Brasileiro de Pesquisas F\'\i
    sicas, Rua Dr. Xavier Sigaud,150, 222\thinspace 90 Rio de Janeiro, Brazil;\qquad \qquad e-mail: baltens@cbpf.br},
  R. Nufer\footnote{Im R\"omergarten 1, CH-4106 Therwil, Switzerland; e-mail: Robert.Nufer@bluewin.ch}}
\maketitle

\abstract{\noindent
\emph{We propose a model for the Pleistocene Ice Age, assuming the following 
scenario: Between 3 Myr and 11.5 kyr BP a Mars-sized object existed which 
moved in a highly eccentric orbit. Originating from this object, gas clouds 
with a complex dynamics reduced Earth's insolation and caused a drop in the 
global temperature. In a close encounter, 11.5 kyr ago, tidal forces deformed 
the Earth. While the shape of the gyroscope Earth relaxed, the North Pole 
moved geographically from Greenland to its present position. During this close 
encounter, the object was torn to pieces, each of which subsequently 
evaporated or plunged into the sun. These events terminated the Ice Age Epoch.}}

\section{Properties of the Plei\-sto\-cene Ice Age} 
Earth's most recent Ice Age Epoch is characterised by unique features, which still require an 
explanation \cite{Elkibbi}. The Pleistocene Glaciation began approximately 2 Myr ago, after a gradual 
decrease of the global temperature in the Upper Plicoene from about 3 to 2 Myr BP. 
During the Pleistocene the general drop in temperature was interrupted by fluctuations, which 
augmented in proportion to the global cooling. Cold periods (Stadials) and warm periods 
(Interstadials) followed each other with a period of about 100 kyr during the last 1 Myr. 
Sometimes the temperature of the Interstadials even exceeded the average value for the 
Holocene \cite{Tiedemann} [Fig.~1]. 
The last Stadial (100 to 11.5 kyr BP) was interrupted about 20 times at irregular 
intervals by sudden temperature increases lasting from a few hundred to a few thousand years 
(Dansgaard-Oeschger events) \cite{Greenland}  [Fig.~2]. During the Last Glacial Maximum, 20 kyr ago, the 
continental ice sheets reached the region around the present New York and covered Northern 
Germany, while Eastern Siberia and part of Alaska were ice-free and inhabited by large 
herbivorous mammals such as mammoths. Some of these have been excavated in a frozen state, 
which shows that in these areas the temperature dropped suddenly at the end of the Ice Age 
Epoch. During the Last Glacial Maximum, the continental ice sheets were centred in a 
geographically displaced position with relation to the present pole positions \cite{Petit-Maire}. 
In the Northern 
Hemisphere this position was in Greenland, about 18 degrees away from the present North Pole 
[Fig.~3]. According to Fig.~2, the Last Glacial Maximum was suddenly terminated 11.5 kyr ago. 
At about the same time a catastrophic geological event occurred, the relics of which are recorded 
in peculiar sediments found all over the Earth \cite{Allan}. 

\begin{figure*}
\includegraphics[width=16cm,keepaspectratio]{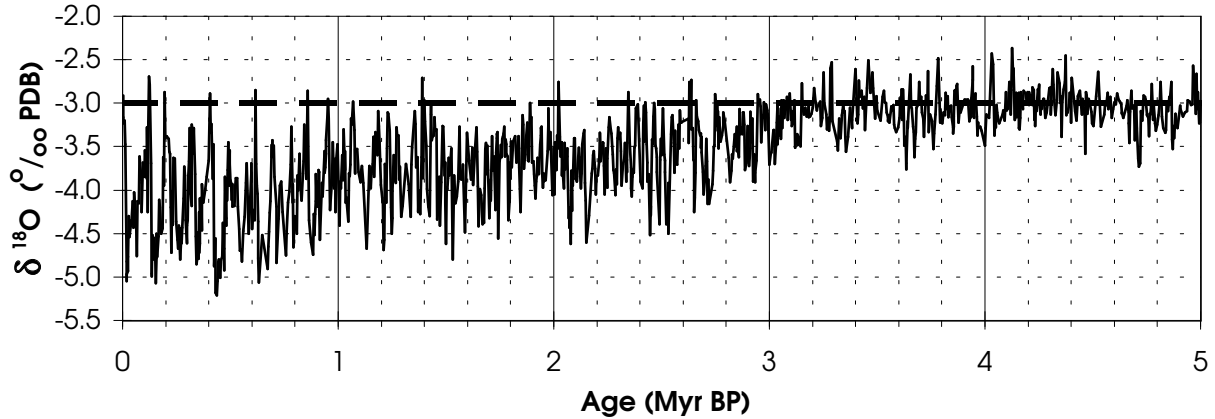}
\caption{Variation of the $\delta^{18}$0 isotope abundance in benthic foraminifera in sea sediments from site 659 
(18$^\circ$ N, 21$^\circ$ W) over the last 5.0 million years \cite{Tiedemann}. Decreasing $\delta^{18}$O corresponds to decreasing 
ice caps and to warmer deep-sea temperature. The late Pliocene (before 3.2 Myr) is characterized 
by remarkably stable and warm climatic conditions comparable to those of the Holocene, the 
average temperature of which is shown by the horizontal dashed line.}
\end{figure*}

\begin{figure*}
\includegraphics[width=16cm,keepaspectratio]{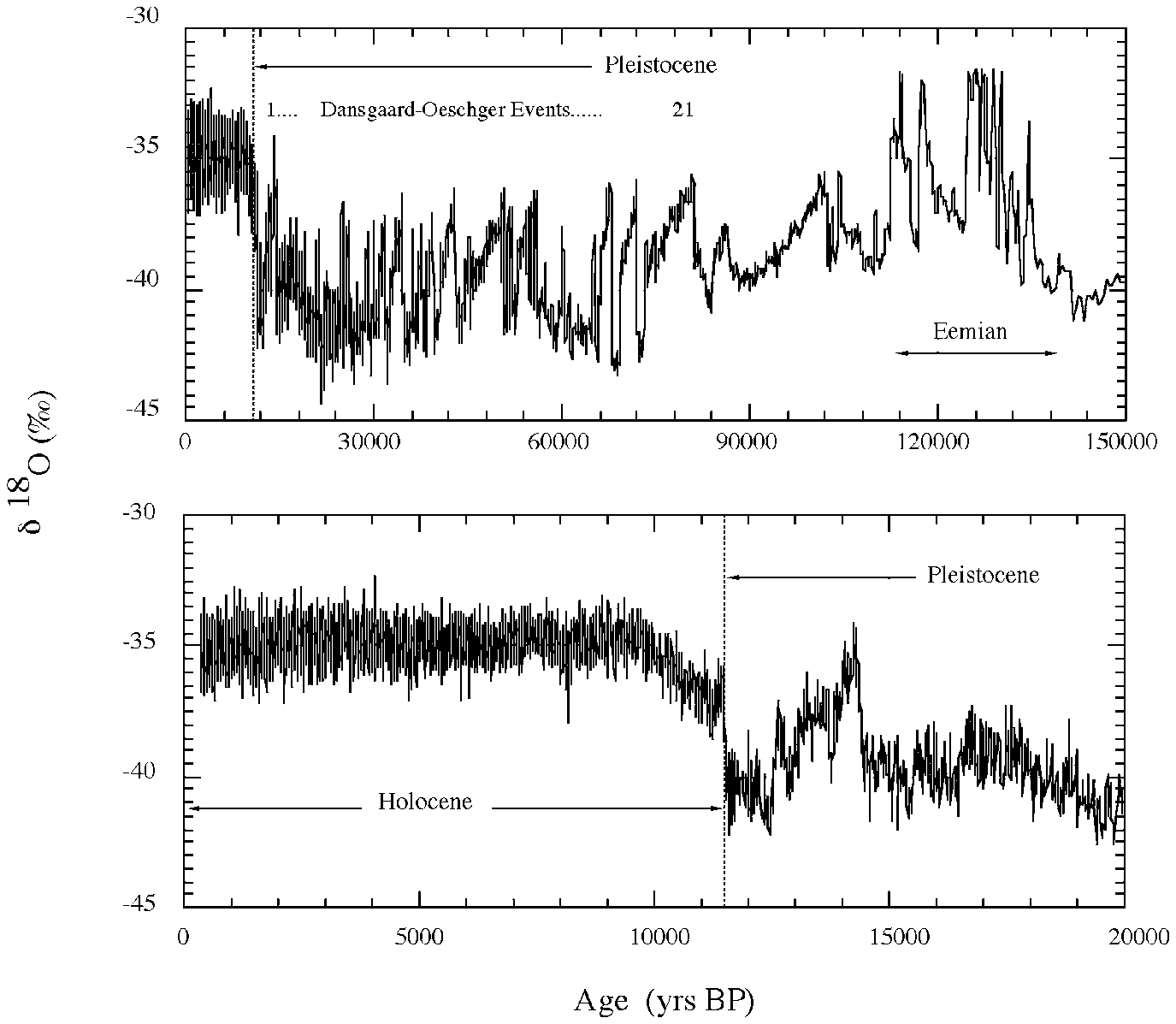}
\caption{The upper part shows the $\delta^{18}$O record of Greenland ice (GRIP) over 150\thinspace 000 years \cite{Greenland}. The 
lower part is an expanded view of the same data over the last 20\thinspace 000 years. Note the difference 
between the last Interstadial, the Eemian, and the Holocene. The last glacial period, 90\thinspace 000 to 
11\thinspace 500 years ago, shows about 20 short temperature variations (Dansgaard-Oeschger events).}
\end{figure*}

\section{An additional planet as the basis of the model}
Usually, the Ice Age Epoch is considered to be the reaction of a highly unstable climate system to 
the slow insolation variations proposed by Milankovitch \cite{Milankovitch}. However, the present climate on 
Earth with its distribution and behaviour follows the basis of the Milankovitch model, suggesting 
that for the Holocene the climate does not require major non-linearities for its explanation. In 
contrast, throughout the Ice Age Epoch, the climate was strongly variable, as is evident from 
Fig.~1. It is therefore reasonable to assume that for a limited time the main driving force of the 
climate was not the Milankovitch effect but an additional external agent to which Earth's climate 
responded linearly. In particular, the asymmetry of the ice distribution in the Northern 
Hemisphere as well as the presence of Mammoths in arctic Siberia during the Last Glacial 
Maximum suggest that the geographic position of the North Pole was located somewhere in 
Central Greenland. If this was the case, then, at the End of the Pleistocene, it had to move to its 
present position. Such a movement of the geographic position of the Earth's rotation axis, which 
in stellar space retains a practically fixed direction, can be induced by a transitory deformation of 
the Earth. This requires an extremely close passage near the Earth of a mass having at least the 
size of Mars. We therefore postulate that during the Ice Age Epoch and in the Upper Pliocene 
such an additional planet existed, henceforth called Z. Since at present Z does not exist any more, 
the Sun is the most likely agent to have promoted its disappearance. We therefore assume that Z 
moved in a highly eccentric orbit with a perihelion distance of only about 4 million km, so that 
during each passage near the Sun, Z was heated by both tidal forces and solar radiation. Thus 
planet Z was liquid and radiant.  

\begin{figure*}


\includegraphics[width=16cm,keepaspectratio]{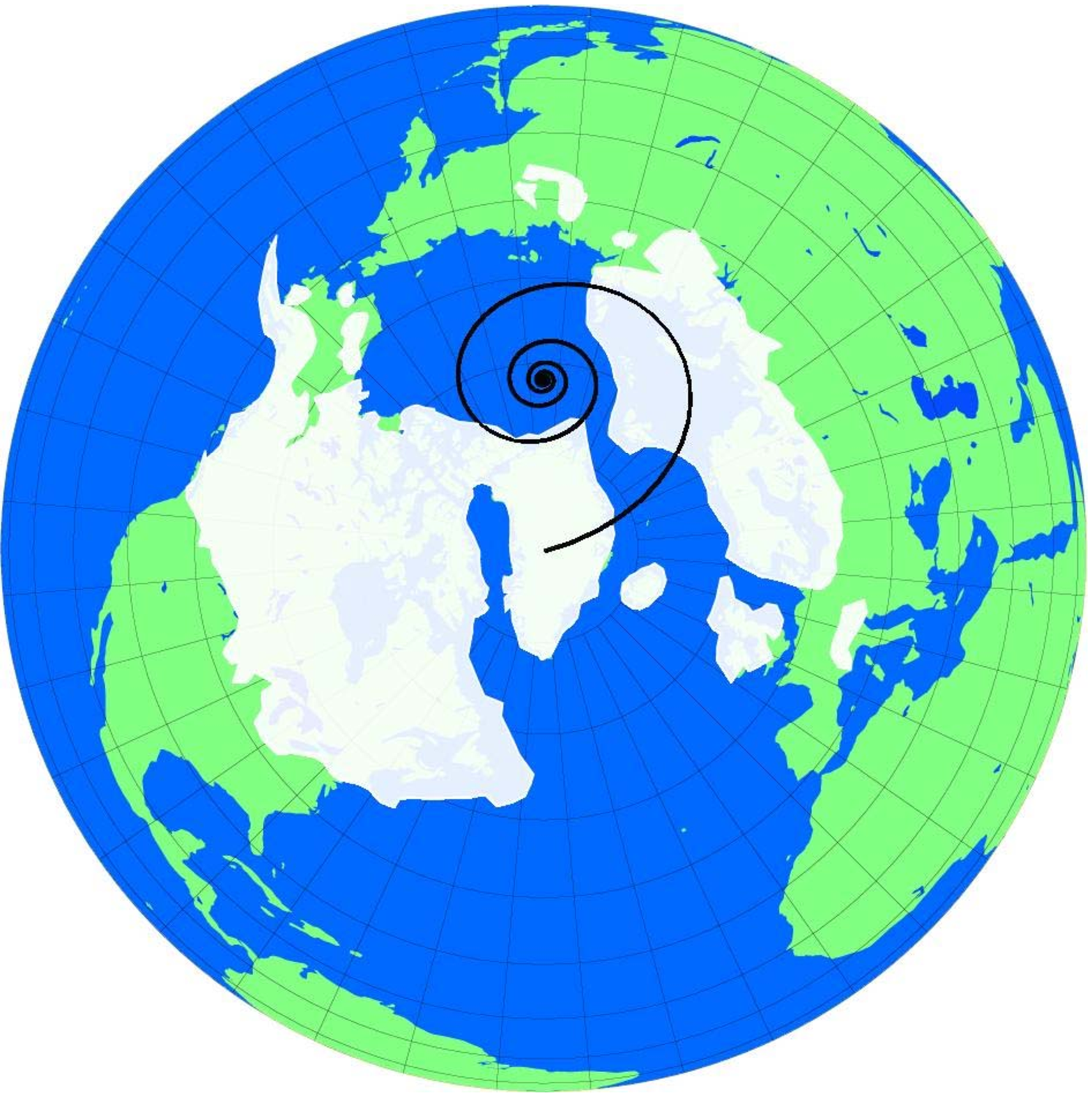}
\caption{The continental ice shield of the Last Glacial Maximum was approximately centred on the central 
part of Greenland \cite{Petit-Maire}.  This suggests that this was the geographic position of the North Pole. 
The spiral shows the geographic migration of the North Pole for a deformed Earth as described in 
the Appendix. Angular momentum is conserved so that in stellar space the rotation axis remains 
practically fixed. The boundary of the permanent ice cover of the Arctic Ocean is uncertain.}
\end{figure*}


\begin{figure*}
\includegraphics[width=16cm,keepaspectratio]{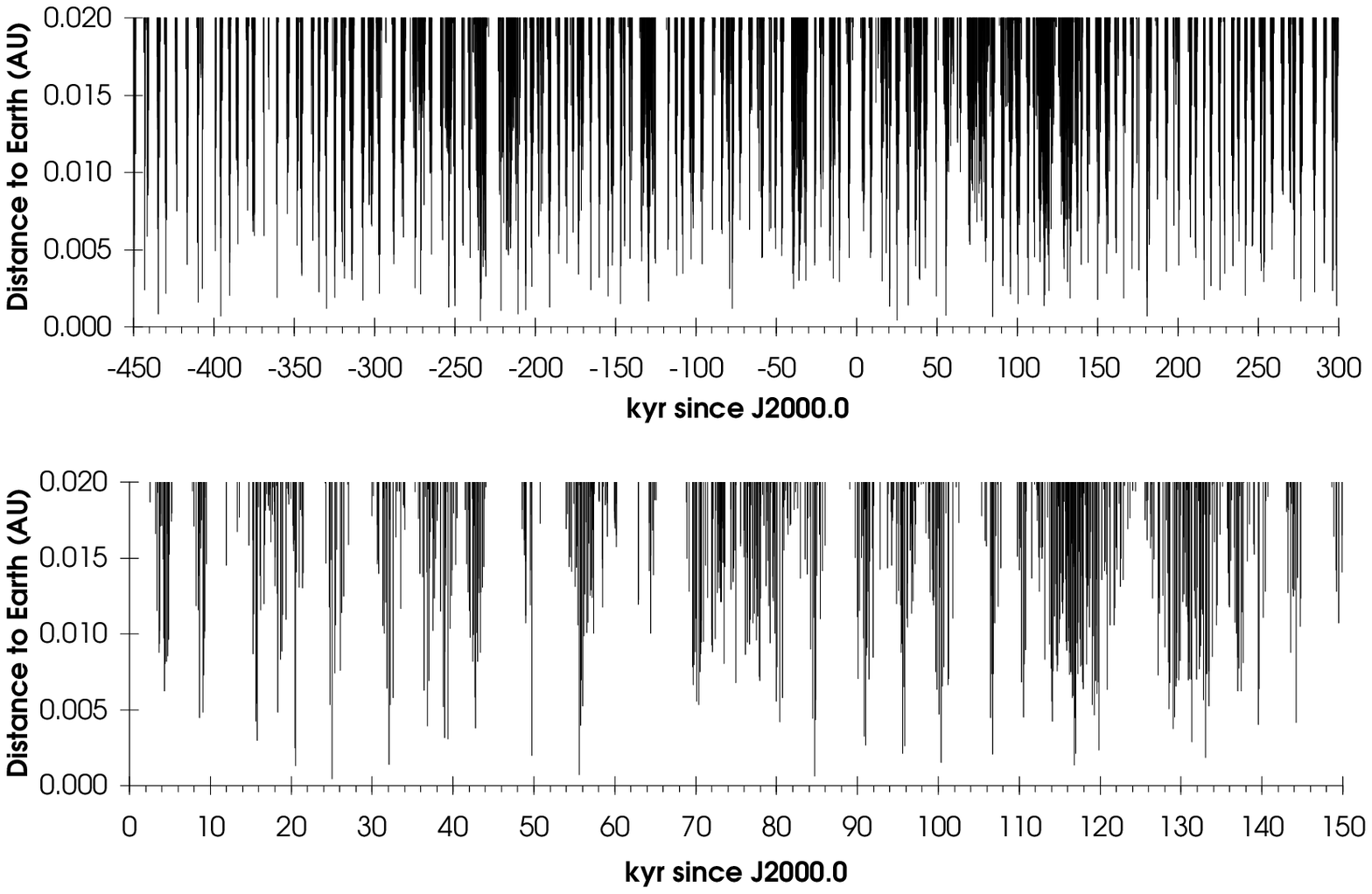}
\caption{Upper panel: Approaches between planet Z and Earth to distances less than 0.02 AU = 3 million 
km as calculated over the past 750\thinspace 000 years. Lower panel: Expanded view of the irregular 
clustering of these approaches over the last 150\thinspace 000 years.} 
\end{figure*}

\section{Origin and fate of Z}
Since Z is not a member of the present planetary system the crucial question has to be answered 
how in such a short time interval it could appear and subsequently disappear. Regarding the 
origin of Z there are several possibilities: Z may have entered into the planetary system from 
outside, i.e. from the Kuiper belt or the Oort cloud. Alternatively, it may have its origin in the 
Asteroid belt or as a moon of Jupiter.  It then must have lost energy and angular momentum 
through resonances with other planets \cite{Murray}. This requires a time of the order of a million years 
only. Fig.~1 suggests that Z reached an orbit with a small perihelion distance 3 million years ago, 
thereby creating a gas cloud resulting in the Earth's Pleistocene.
	Regarding the termination of Z, it is indispensable to assume that it was fragmented 
during the final close encounter with the Earth.  This process consumed orbital energy, so that the 
perihelion distances of the fragments were likely to be reduced compared to the perihelion 
distance of Z. Most importantly, the smaller escape velocity of the fragments increased the 
evaporation rate. Typically, the molecular binding energy became more important than the escape 
energy, so that both molecules and clusters could evaporate. These were then blown away by 
radiation pressure. In this way, the fragments of Z could become dissolved within the Holocene. 
It is not unlikely that some of the fragments also dropped into the Sun. 

\begin{figure}
\includegraphics[width=7.6cm,keepaspectratio]{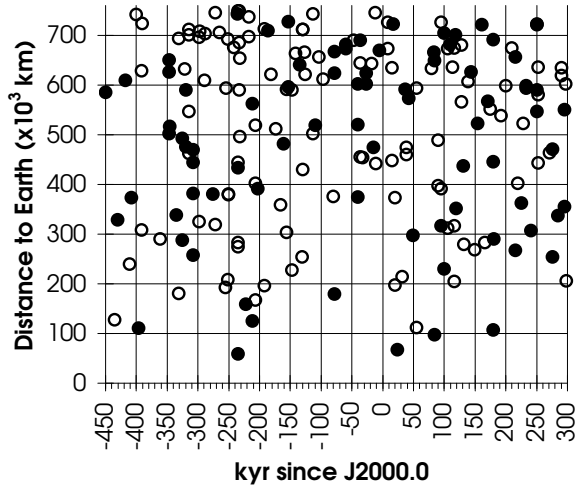}
\caption{Closest approaches over 750\thinspace 000 years below twice Moon-Earth distance. The two horizontal 
dashed lines indicate, respectively, Moon's distance (384\thinspace 000 km) and the distance below which 
significant polar shifts are to be expected (30\thinspace 000 km). The open (filled) circles mark encounters 
during which Z moves away from (towards) the Sun.}
\end{figure}

\section{Frequency of approaches to Earth}
In order to obtain a measure of how often Z approached the Earth, the equations of motion of the 
planetary system including Z were solved for various orbital parameters of Z. Tidal work and any 
effects from evaporation were disregarded.  Earth and Moon were considered separately, and the 
planets beyond Saturn were omitted \cite{Nufer}. In the main calculation, the parameters assumed for Z at 
time J2000.0 were: semi-major axis 0.978 AU, numerical eccentricity 0.973, inclination 0$^\circ$, 
longitude and argument of the perihelion both 0$^\circ$. The result for a period of 750 kyr is shown in 
Figs. 4 and 5.  The orbit of Z was found to be stable over the time range considered. The semi-
major axis varied within 0.95 AU and 1.1 AU without showing a general trend. Similarly, the 
eccentricity remained between 0.958 and 0.977. The inclination, i.e. the angle between the orbit 
of Z and the invariant plane, showed an irregular variation between 0$^\circ$ and 13.5$^\circ$, with a period of 
approximately 9 kyr.  In Fig.~4, each approach to Earth to less than 3 million km is marked by a 
vertical line ending at the closest distance. The Figure shows that the encounters are irregularly 
clustered, somewhat resembling the pattern of temperature fluctuations of Figs. 1 and 2. In Fig.~5 
calculated approaches between Z and Earth to less than twice the Earth-Moon distance are 
shown. Within 100 kyr there are several passages closer than Moon's distance. These must have 
created enormous earthquakes, and during the Ice Age Epoch must have caused ruptures of the 
continental ice shelves. We tentatively identify these with the Heinrich events \cite{Heinrich}. The diagram in 
Fig.~5 contains no approach closer than 30 000 km, i.e., approaches which might induce a polar 
shift; however, additional calculations suggest that this may occur once in a few Myr.

\section{Mechanics of a polar shift}
The asymmetry of the glaciation in the Northern Hemisphere [Fig.~3] is the most conspicuous 
feature of the Ice Age Epoch. Because of this, a displaced pole position and a fast polar migration 
have been postulated already at the end of the 19$^{\hbox{\scriptsize{th}}}$ century \cite{Hapgood}. This migration is a damped 
precession of the rotation axis on the globe, while the axis remains fixed in stellar space. This 
was discussed and judged as impossible. Indeed, with an Earth in either the solid or the liquid 
state, both of which were at that time considered, a deformation would relax too fast to allow an 
appreciable geographic shift of the poles. However, assuming a plastic Earth with a relaxation 
time of at least a few hundred days makes possible a shift of the required magnitude of ca. 18$^\circ$ 
\cite{Gold}.  
The rotation of the unperturbed Earth is stabilised by its increased radius at the equator 
compared to that at the poles. If the Earth had an additional deformation in an oblique direction, it 
would perform a motion in which the position of the rotation axis on the globe would migrate. 
The deformation could result from a close passage of a planet-sized object, which would stretch 
the Earth by tidal forces. A 1 per mil deformation is required for the shift. We consider an initial 
stretching deformation of the Earth such that, at an angle of 30$^\circ$ with the initial rotation axis, the 
radius is increased by 6.5 km. In the ensuing process, the Earth relaxes into a new equilibrium 
shape with a displaced equatorial belt. Angular momentum is strictly conserved, and at all times 
the rotation axis practically points to the same star. The spiral in Fig.~3 shows the geographic path 
of the rotation axis. It is the solution of Euler's equations and a relaxation equation for the inertial 
tensor. A turn in the spiral takes about 400 days.  For details of the calculation, see Appendix and 
\cite{Woelfli}.
A stretching force is obtained from the tidal action due to a mass near the Earth. The 
required deformation corresponds to the equilibrium shape of Earth when a Mars-sized mass is at 
24\thinspace 000 km distance. The actual deformation problem, with a mass passing near the rotating 
Earth, is vastly more complex. We expect that the peak tidal force has to be about an order of 
magnitude larger; this brings the closest distance between the centres of Earth and Z into the 
range of 12\thinspace 000 to 15\thinspace 000 km. As a result, Z enters the Roche limit of Earth. It is then likely to be 
torn into two or more parts.  Since Z is lighter than the Earth, tidal effects are stronger on Z than 
on Earth: Z is torn to pieces rather than the Earth. Note that the requirements restrict the mass of 
Z at the end of the Pleistocene to a range of at least that of Mars and clearly smaller than Earth's. 
The polar shift event must have been accompanied on Earth by continental floods, earthquakes 
and volcanic eruptions, i.e. a world-wide catastrophe, which actually left many stratigraphic 
evidences and which can also be assumed to be reported in traditions in many countries all over 
the globe \cite{Allan}.

\section{The gas cloud during the Ice Age Epoch}
Since during the Pleistocene Z is assumed to have been at least Mars-sized, evaporation was 
limited by the gravitational escape energy. Only single atoms escaped from the hot surface of Z. 
If an atom has an optical transition from the ground state within the main solar spectrum, then 
radiation pressure expels it from the planetary system.  However, some atoms and many ions can 
be excited with ultraviolet light only. Apart from the rare gas atoms, these include atoms of 
Oxygen and Carbon. In these cases the repulsion due to solar radiation is much weaker than the 
gravitational attraction to the Sun, so that atoms of these elements can remain in bound orbits. 
These orbits shrink due to the Poynting-Robertson drag.
	The continued evaporation creates an interplanetary cloud. Its material consists of single 
atoms and ions and is thus quite distinct from the present zodiacal dust \cite{Gustafson}. The dynamics of the 
atomic cloud involves a variety of processes and is complex. We can only tentatively guess its 
behaviour. Collisions between atoms with planetary velocities are inelastic. They reduce the 
relative velocity between the colliding particles, so that their outgoing orbits become more 
similar. This increases the particle densities and thereby the frequency of collisions. This suggests 
that the range of inclinations of the orbits in the cloud can shrink with a time scale determined by 
the mean free time for particle collisions. Also, since collisions are inelastic, the semi-axes 
diminish. Particles with different ratios of repulsion by solar radiation to gravitational attraction 
intrinsically belong to different orbits and may become separated. If molecules form, these are 
expelled by radiation pressure. Atoms and molecules may become ionised.  
The scattering of solar radiation from any material along the line between Sun and Earth 
lowers the global temperature. Clearly, this screening depends on the density of the cloud and on 
the relative motions of cloud and Earth. Therefore, the extremely strong variations in temperature 
characteristic of the Pleistocene may be due to changes in the screening of the Sun. 
The isotopic and stratigraphic data for the last Myr of the Ice Age Epoch show a 100 kyr 
period \cite{Petit}. Now, Earth's inclination, i.e. the angle between Earth's orbit and the invariant plane, 
is governed by a 100 kyr cycle. The maxima of the inclination in fact coincide with the 
Interstadials, except for the last maximum, where no Interstadial has been observed. This 
indicates that the orbits of the atoms of the cloud often had inclinations which were smaller than 
the maxima of Earth's inclination. Possibly at the end of the Pleistocene the width of the cloud 
had become too large. A similar solution to the problem of the origin of the 100 kyr cycle has 
previously been suggested by Muller et al \cite{Muller95,Muller97}.

\section{The termination of the Ice Age Epoch}
	The climatic fluctuations which occurred towards the end of the Pleistocene require 
further study. We just note here that the last rapid increase of the temperature recorded in the 
polar ice data occurred at 11\thinspace 500$\pm$65 yr BP [Fig.~2]. All radiocarbon dates made on residue 
material originated during the global catastrophe point to the same age \cite{Allan,Martin}). However, these 
radiocarbon ages are not corrected for the variation of the production rate. The new dendro and 
U/Th calibration curves indicate that these ages have to be increased by about 1500 yr \cite{Stuiver,Bard}). 
Thus, it appears that the Younger Dryas, which begins at 12\thinspace 700$\pm$100 yr BP \cite{Greenland} is younger than 
the polar shift event. At the beginning of the Holocene the temperature increased in two steps. 
The first fast step was followed by a much slower rise, which reached its maximum about 9\thinspace 000 
yr ago [Fig.~2]. Since then the temperature remained remarkably constant until today. The 
possibility should not be a priori discarded that minor, still unexplained climatic features such as 
the cold events at 8.2 kyr cal BP \cite{Hu} and 4.166 kyr cal BP \cite{Beck} as well as the so-called Little Ice 
Age, 300 years ago, are due to remaining traces of gaseous material.

\section{Facts which become plausible or which can be explained by our model}

\subsection{From the displaced pole positions} 
\begin{itemize}
\item{The {\bf asymmetry} of the glaciation in North America and east Siberia; it was the main 
motivation for considering geographically displaced poles. At present there are no climate 
models which determine the optimum position of the poles and the amount of screening 
of the solar radiation compatible with the observed glaciation.} 
\item{The existence of {\bf mammoths} in arctic East Siberia indicates that there was sufficient 
sunlight for the growth of the plants on which they lived.}
\item{{\bf Archaeological} objects having ages around 40\thinspace 000 years BP were found close to the 
Arctic Circle in Siberia  \cite{Pavlov}. At the time the place had lower latitude.}  
\item{{\bf Lake Baikal} was never frozen during the Pleistocene \cite{Kashiwaya}.}
\item{The {\bf Tibetan Plateau} was about 15$^\circ$ closer to the equator and not ice-domed during the last 170 kyr \cite{Schaefer}.}
\item{The {\bf Atacama Desert} and the {\bf altiplano Bolivia} were humid \cite{Baker}. They were then situated 
closer to the equator than today.}
\item{The {\bf Sahara desert} was covered with grass and bushes during the Pleistocene. While for 
the Western Sahara this may be explained by its higher latitude in the Pleistocene, other 
reasons such as globally lower temperature should play a role in the Eastern Sahara.} 
\end{itemize}

\subsection{From the interplanetary gas cloud} 
\begin{itemize}
\item{The {\bf beginning} of the general temperature decrease, 3 Myr ago can be understood as the 
time at which the perihelion distance of Z became sufficiently small for the gas cloud to 
develop.}
\item{The {\bf coldest} Ice Age was at the {\bf end} of the Pleistocene, since due to tidal work the 
perihelion distance of Z decreased. With the passing of time the cloud became denser.}
\item{{\bf The colder} the mean temperature, {\bf the larger} were the {\bf variations} of the temperature.  
Dense clouds throw strong shadows.} 
\item{Some Interstadials had {\bf higher} temperatures than at present. Not only was the Earth 
exposed to the regular sunlight, but it also received radiation scattered from clouds or 
backscattered from material outside Earth's orbit.} 
\item{The form of the {\bf Daansgard-Oeschger temperature variations}, typically a gradual 
decrease followed by a rapid increase, may be connected with the presently still unknown 
dynamics of the cloud. The observed shape of the variations may help to understand the 
cloud's behavior.} 
\item{During the last million years the Interstadials occur approximately every {\bf 100 kyr}, which 
coincides with  the cycle of Earth's inclination.}
\end{itemize}

\subsection{From the orbit of Z}
\begin{itemize}
\item{The approaches shown in Fig.~5 produce a similar irregular pattern as the temperature 
variations of Fig.~2.  The events marked in Fig.~5 have distances less than 3 Mio. km, 
which is the radius to which a gas cloud may expand by thermal motion during the 
passage from the perihelion of Z to Earth's distance. This suggests that a screening of the 
solar radiation by gases may lead to a {\bf frequency of temperature excursions} as observed in the 
sequence of the Dansgaard-Oeschger events.} 
\item{Gigantic earthquakes, as listed in Fig.~5, accompanied fly-byes at less than Moon's 
distance. Their frequency, about 6 per 100 kyr, corresponds to the {\bf frequency of Heinrich 
events}, in which large glaciers broke away from the continent and floated into the 
Atlantic Ocean.}
\item{Fig.~5 contains no fly-byes to less than 30\thinspace 000 km, which might induce a polar shift. The 
{\bf rarity of polar shifts}, say one event in a few Myr, is compatible with Fig.~5.}
\end{itemize}

\subsection{From the polar shift catastrophe}
\begin{itemize}
\item{The model explains how a {\bf polar shift} within the time of relaxation of a global 
deformation, i.e. a few years, can occur.}
\item{The {\bf frozen mammoths} in the permafrost of arctic Siberia are a direct consequence of the 
geographic motion of the polar axis.}  
\item{The catastrophe produced a global {\bf extinction of} many {\bf species} of large animals \cite{Allan,Martin}.}  
\item{ {\bf Frozen muck} containing broken trees and bones testify for the violence of the event \cite{Allan}.} 
\item{The fragmentation of Z lead to its relatively rapid disappearance. Thus the Ice Age Epoch 
had an {\bf end}.}
\item{Once the continental ice shelf was molten, the {\bf Holocene} was {\bf constantly warm} and 
distinct from the Interstadials, which were interrupted by temperature variations.}  
\item{Human {\bf civilization reappeared} about 9\thinspace 000 years ago. Notably these populations, which 
stem from survivors, had an elaborately structured language.} 
\item{Many {\bf traditions} are related to Z or its fragments, such as the Chinese dragon, a flying 
animal that spits fire and has a long, indefinite tail \cite{Velikovsky}.} 
\end{itemize}

\section{Conclusion}
At present, the climate system of the Earth is observed to react to external forcing in a plausible 
way. If we assume this to be generally true in the geological past the observed asymmetry of the 
glaciation during the Late Glacial Maximum requires a shifted pole position and a fast migration 
of the poles at the end of the last glaciation. A pole shift of the order of 18$^\circ$ requires a close 
encounter with a massive object, which we have here called Z. Its mass was at least Mars sized, 
but clearly smaller than Earth's, so that Z could be torn into fragments during the encounter. 
Since only the Sun can dispose of Z, we have to assume the perihelion of Z so small that Z was 
intensely heated. Clearly, its aphelion has to be larger than the radius of the Earth's orbit. The 
choice of the aphelion determines the frequency of encounters. It had several million passes near 
the sun and a few close encounters with the Earth as well as with the other inner Planets. The 
heating of the surface of Z lead to the accumulation of an interplanetary atomic gas cloud having  
a complex dynamics. The material between the Sun and the Earth reduced the insolation and 
thereby the global temperature in a time-dependent way. In particular, the increasingly cold and 
variable climate during the last 3 Myr, until 11.5 kyr ago, is plausibly explained by the slow 
decrease of the orbital energy of Z and its angular momentum. The perihelion decreases, and this 
enhances the density of the gas cloud. In the Holocene, after the removal of all fragments of Z, its 
threat for life on Earth finally ended. Note that this model has only few free parameters. On the 
other hand, it creates new problems that deserve a more complete treatment in future studies.
It may be worthwhile to clarify the relation of our model to a claim held by I. Velikovsky 
\cite{Velikovsky} that a close-by passage of Venus and later of Mars had produced a polar shift on Earth. 
Einstein \cite{Einstein} in his third letter to Velikovsky resumed his recommendations by the expression 
"Catastrophes yes, Venus no". Our model is compatible with this directive of Einstein. 

\section*{Appendix} 
The tidal force field $F(z)$ (for large distances $R$ compared to the radius $R_E$  of the Earth) is parallel to the $z$-direction, which points to the perturbing mass $M_Z$ . It has the value
\begin{equation}
F(z) = 2 M_Z G\; \frac z{R^3}, 
\end{equation}
where $G = 6.673\cdot 10^{-11}$  m$^3$ kg$^{-1}$ s$^{-2}$   is the gravitational constant.  Earth's induced deformation is described by an increment to the radius $H(\gamma )$, where $\gamma$ is the angle with the direction $z$ (at latitude 30$^\circ$)
\begin{equation}
H(\gamma ) = H_0 \left[\cos(2\gamma )+\frac 13\right] .            	                   				
\end{equation}
The energy for such a deformation is minimised by the amplitude
\begin{equation}
H_0 = \frac{R_E^2 M_Z G}{2 R^3 g}							
\end{equation}
with $g = 9.8$ m/s$^2$, the gravitational acceleration at the surface of the Earth. For $R = $24\thinspace 000 km, $H_0  = 6.45$ km. In a dynamic theory $R$ will be smaller.
The diagonalized inertial tensor of the equilibrium Earth $\Xi_0$ has matrix elements [1.0033,1,1] in units $8.01\cdot 10^{37}$ kg m$^2$. For the deformed Earth the initial inertial tensor $\Xi(0)$ has diagonal elements [1.0018, 0.9995, 1] and off-diagonal elements $\Xi_{12}(0) = \Xi_{21}(0) = 0.000\thinspace 9$.  Deformations are assumed to relax as
\begin{equation}
\frac{d \Xi}{dt} = -\frac{\Xi(t) - \Xi_0[\vec\omega (t)]}\tau ,\qquad		\tau = 1000\; \hbox{d}.			
\end{equation}
The geographic wandering of the rotation vector $\vec\omega$ in co-ordinates fixed to the Earth is described by the Euler equation for free motion:
\begin{equation}
\frac{d \;\Xi \vec\omega}{dt} = [\Xi\vec\omega,\vec\omega].							
\end{equation}                          
Eq. (4) and (5) are solved numerically.

\section*{Acknowledgement} 
We are indebted to H.-U. Nissen for comments on the manuscript.

\end{document}